\documentclass[twocolumn,showpacs,preprintnumbers,amsmath,amssymb,prl]{revtex4}

\usepackage{graphicx}
\usepackage{dcolumn}
\usepackage{epsf}

\begin{document}

\title{Thickness dependence of critical currents in thin superconductors}
\author{A. Gurevich}
\affiliation{Applied Superconductivity Center, University of Wisconsin, Madison, Wisconsin 53706}

\date{\today}

\begin{abstract}
Mechanisms of vortex dynamics and pinning and self-field effects which could account for the  
thickness dependence of the critical current density $J_c$ of superconducting 
films are addressed. It is shown that at low magnetic fields B, the 2D single-vortex pinning   
and thermal fluctuations yield $J_c\propto (1/\sqrt{d})\exp[ - (d_m/d)^{1/2}]$, if the film thickness d is smaller than the 
pinning correlation length along B. The model describes the dependence of $J_c$ on d observed 
on $YBa_2Cu_3O_7$ coated conductors for $d<2\mu$m. Measurements of $J_c(d)$ in strong magnetic fields 
are proposed to probe whether the dependence of $J_c$ on d is mostly determined by the 2D pinning or  
changes in the material microstructure as the film gets thicker.  

\end{abstract}
]

\maketitle

Critical current densities $J_c$ of superconducting films are typically much greater (by orders of magnitude) than 
those of bulk conductors. This difference is usually ascribed to denser and stronger pinning defect microstructures 
in epitaxial films, lesser effect of current-blocking obstacles such as grain boundaries, microcracks, etc. A general 
trend is that $J_c(d)$ first increases as the film thickness $d$ decreases, followed by a decrease of 
$J_c$ and the critical temperature $T_c$ for smaller $d$. This issue of what is behind the size dependence of $J_c$
is important for high-$T_c$ multifilamentary conductors and $YBa_2Cu_3O_7$ (YBCO) coated conductors, in which 
a significant degradation of $J_c(d)$ as the YBCO film thickness increases was reported \cite{lanl1,lanl2,ornl}.
This behavior of $J_c(d)$ has been attributed to the change of the microstructure as the YBCO film gets thicker, 
in particular,  "dead" layers with reduced $T_c$ and $J_c$ near the buffer layer and the surface, lesser epitaxy and the 
biaxial texture, etc. \cite{lanl1,lanl2,ornl,nature}   

In this Letter I propose an alternative approach, based on the fact that pinning of vortices of finite length 
depends on the film thickness if it gets smaller than the bulk pinning correlation length $L_c$ along 
the field direction. For bulk superconductors, $J_c$ is determined by the balance of the Lorentz 
force $JBV_c/c$ and the pinning force $(n_if_p^2V_c)^{1/2}$ in the correlation volume $V_c=R_c^2L_c$, 
where $R_c$ is the correlation length transverse to 
the field direction, $n_i$ is the volume density of pinning centers, and $f_p$ is an 
elementary pinning force per pin \cite{blat}. If $d$ becomes smaller than $2L_c$, a crossover occurs from the 3D 
thickness-independent bulk pinning to the 2D collective pinning for which $J_c(d)$ depends on d \cite{2d,kes,brandt,moge}. 
This transition was indeed observed on $Nb_3Ge$, $Mo_3Si$ \cite{2d,kes} and $Mo_xGe_{1-x}$ \cite{moge} films.  

For a film of thickness d and width w in a perpendicular magnetic field, the 2D collective pinning theory gives \cite{brandt}
	\begin{eqnarray}
	J_c=\frac{c\gamma}{Br_p^3C_{66}d}[\ln(2l_c/R_c)/8\pi]^{1/2}, 
	\label{ehb1}\\
	R_c=r_p^2C_{66}[8\pi d/\gamma\ln(2a/R_c)]^{1/2}.
	\label{ehb2}
	\end{eqnarray}
Here $R_c$ is the pinning correlation length transverse to the vortex lines, 
$C_{66}\simeq\phi_0B(1-b)^2/64\pi^2\lambda^2$ is the shear elastic modulus, $\lambda$ is the London penetration depth, 
$l_c=(C_{66}cr_p/BJ_c)^{1/2}$, 
$\phi_0$ is the flux quantum, $b=B/B_{c2}$, $B_{c2}$ is the upper critical field, and c is the speed of light. 
The parameter $\gamma = f_p^2r_p^2n_i$ quantifies the pinning strength of  
randomly distributed point defects, each acting on the vortex with the maximum force $f_p$ in the 
interaction radius $r_p$. As follows from Eq. (\ref{ehb1}), the critical current $I_c=J_cdw$ is nearly independent of $d$. 
The condition of the 2D pinning, $d<2L_c$, has the form \cite{2d} 
	\begin{equation}
	d [\mu m] < 2 B^{1/4}[Tesla]/\sqrt{J_c}[MA/cm^2]
	\label{td} 
	\end{equation}  
For $B=1T$ and $J_c=1MA/cm^2$, the transition from the thickness-independent 3D pinning $J_c$ to the 
2D pinning $J_c\propto 1/d$ occurs for $d$ below $\sim 2\mu$m, in qualitative 
agreement with the experimental data on coated conductors \cite{lanl1,lanl2}.  
Thus, the dependence $J_c(d)$ observed on coated conductors may be due to the crossover from the 3D to 2D collective 
pinning, even if the pinning microstructure is unaffected by the film thickness. 

Eqs. (\ref{ehb1}) and (\ref{ehb2}) correspond to high fields, for which $R_c$ is much greater 
than the intervortex spacing $a=(\phi_0/B)^{1/2}$, and pinning-induced deformations of the vortex 
lattice are described by the continuum elasticity theory \cite{brandt}. The condition $R>a$ is equivalent to $B>B_s$, 
where the crossover field $B_s\ll B_{c2}$ can be found from Eqs. (\ref{ehb1}) and (\ref{ehb2}):
	\begin{equation}
	B_s\simeq (8\pi)^{3/2}J_c\lambda^2/cr_p\ln^{1/2}(2l_c/R).
	\label{bs}
	\end{equation}
If $r_p$ equals the coherence length $\xi$ (core pinning),  
$J_c=1 MA/cm^2$, $\lambda(77K)\simeq 0.6 \mu$m, $\lambda/\xi=100$, and $\ln(2l_c/R)\approx 1$, 
Eq. (\ref{bs}) yields the crossover field $B_s\simeq 7.6$ T much higher than the typical field range 
in which most of the $J_c$ measurements on coated conductors have been performed. 
So we turn to the 2D single-vortex pinning at low fields $B<B_s$, for which the macroscopic elasticity of the vortex lattice is irrelevant. 

We first consider self-field limitations of $J_c$ at zero field, and estimate a mean current density 
$J_{c1}=cB_{c1}/2\pi d$ which produces the parallel field equal to the lower critical field 
$B_{c1}=(\phi_0/4\pi\lambda_{ab}\lambda_c)[\ln(\lambda/\xi)+0.5]$ in the middle of the film. For typical YBCO parameters, 
$\lambda_{ab}(77)=0.6 \mu$m, $\lambda_c/\lambda_{ab}=7$, $d = 2 \mu$m,    
$B_{c1}(77K)=14.7$ Oe, the threshold value $J_{c1}$ is about $0.12 MA/cm^2$.
The so-obtained lower critical current density $J_{c1}$ is an order of magnitude 
lower than typical $J_c$ values of coated conductors \cite{lanl1,lanl2}, so the onset of dissipation 
at $J_c$ is most likely due to depinning of self-field vortices which penetrated the film at $J\ll J_c$. 
However, the onset of flux penetration and the $J_{c1}$ values are strongly affected by the surface and geometrical 
barriers which can be locally reduced by surface and edge defects in the YBCO film and buffer layer. 
Thus, $J_{c1}$ values can vary in a wide range $cB_{c1}/2\pi d < J_{c1} < cB_{c}/2\pi d$, 
where the upper limit corresponds to an ideal surface, and $B_c=\phi_0/2\sqrt{2}\pi\xi\lambda$ is the 
thermodynamic critical field. The self-field effects could therefore contribute to the size dependence of $J_c$ 
in zero field, but they disappear at higher fields much greater than the full penetration 
field $B_p\simeq J_cd/c$.

Now we consider the 2D single-vortex pinning in a film in a perpendicular field 
$B < B_s$. For $d<L$, pinning only causes overall displacements, but not distortions of vortex lines which thus 
remain nearly straight and perpendicular to the film surface. In this case even a rigid vortex of finite length $d$ is pinned by 
randomly distributed point defects, unlike an infinite sample where 
the pinning force of a straight vortex in a random pinning potential is zero. Indeed, let the film thickness be much greater than the 
mean spacing between the pinning centers, then the vortex interacts with $N=r_p^2dn_i$ pinning centers,  
adjusting its position so that the net pinning force vanishes (see inset in Fig. 1). The transport current 
displaces the vortex from the local minimum of the pinning potential, so now the random elementary pinning forces 
do not exactly compensate each others. The balance of the total maximum pinning force $f_p\sqrt{N}$ 
from uncorrelated pinning centers in a {\it finite} volume $r_p^2d$ and the Lorentz force $\phi_0dJ/c$ yields the 
following equation for $J_c$:
	\begin{equation}
	\phi_0J_cd/c=f_p\sqrt{N}\exp(-u^2/r_p^2).
	\label{bal}
	\end{equation}
Here the exponential Debye-Waller factor accounts for the 
smearing of the pinning force by thermal vortex fluctuations \cite{blat}.
The mean-squared vortex displacement $u^2$ is given by the equipartition theorem:
	\begin{equation}
	\alpha_Lu^2 = kT,\qquad\qquad \alpha_L=f_p\sqrt{N}/r_p,
	\label{eq}
	\end{equation}  
where k is the Boltzmann constant, and $\alpha_L$ is the Labush 
spring constant. Eqs. (\ref{bal}) and (\ref{eq}) yield 
	\begin{equation}
	J_c=\frac{c}{\phi_0}\sqrt{\frac{\gamma}{d}}\exp\bigl(-\frac{kT}{r_p\sqrt{\gamma d}}\bigr),
	\label{jc}
	\end{equation} 
where $\gamma (T,B)=U_p^2n_i$, and $U_p=f_pr_p$ is the elementary pinning energy. The function $J_c(d)$ shown 
in Fig. 1 has a maximum at $d=d_m$ due to the competition of pinning and thermal fluctuations. 
Indeed, pinning increases $J_c\propto 1/\sqrt{d}$ due to enhancement of relative fluctuations of the pinning potential 
$\sim N^{-1/2}$ as d decreases. At the same time, the amplitude of thermal vortex vibrations $u\propto d^{-1/4}$ in the 
exponential factor also increases as d decreases. Here $d_m$ and the maximum current density $J_m=J_c(d_m)$ are given by	
	\begin{equation}
	d_m=\frac{k^2T^2}{\gamma r_p^2}, \qquad\qquad J_m=\frac{c\gamma r_p}{e\phi_0 kT},
	\label{jm}
	\end{equation}
where $e=2.718$. The optimum thickness $d_m(T)$ increases at $T$ increases, diverging at $T_c$. 
For instance, for randomly distributed point pinning centers (say, oxygen vacancies \cite{beasley,dcl}),  
$r_p\simeq \xi(T)$, and $U_p\sim B_c^2(1-b)^2\sigma\xi(0)$ and $U_p\sim B_c^2(1-b)^2\sigma\xi^2/\xi(0)$ for the s and 
d-wave pairing, respectively, where $\sigma$ is the scattering cross-section of the pin \cite{pin}. Hence, 
$\gamma(T)\simeq \gamma_0[1-(T/T_c)^2]^2$ for the d-wave high-$T_c$ superconductors. The value $d_m$ can also 
be written in the form independent of particular pinning mechanisms:
	\begin{equation}
	d_m = \bigl[ ckT/J_{c}(d_1,T,B)\phi_0r_p\sqrt{d_1}\bigr]^2,
	\label{dm}
	\end{equation}
where the parameter $\gamma=d_1(J_c\phi_0/c)^2$ is expressed in terms of the observed 
$J_c$ at a sufficiently large thickness $d_1>d_m$ for which the 2D pinning still holds, but the 
thermal exponential factor in Eq. (\ref{jc}) is close to 1. For $J_{c}= 1MA/cm^2$ at $d_1=1 \mu m$ and 
$T=77 K$ \cite{lanl2}, $r_p=\xi(77K)=60 \AA$, Eq. (\ref{dm}) yields $d_m\simeq 0.8 \AA$. This small values 
of $d_m$ indicates that thermal fluctuations do not affect $J_c$ of YBCO coated conductors 
at $77$ K and B well below the irreversibility field $B^*$. However, the situation changes for low-$T_c$ weak pinning 
superconductors with $J_c\sim 10^3-10^4 A/cm^2$, and $d\sim 0.02-1 \mu$m, such as $Nb_3Ge$, $Mo_3Si$ \cite{kes} 
and $Mo_xGe_{1-x}$ \cite{beasley} films. For instance, for $Mo_xGe_{1-x}$ amorphous films with $J_c=5\times10^3 A/cm^2$, 
$T_c=7.2$ K, and $d_1=1200 \AA$, Eq. (\ref{dm}) yields $d_m=780 \AA$ for $r_p=\xi(T)=60 \AA$ and $T=4.2$ K. Because $d_m$ 
strongly depends on T and B, the appropriate tuning of T and B can bring $d_m$ into the range of 
$d\sim 10^2-10^4 \AA$ for which the nonmonotonic dependence of $J_c(d)$ can be observed. Thus, low-$T_c$ 
films can be convenient model systems to probe the 2D-3D pinning crossover, and the interplay between vortex 
pinning and thermal fluctuations, which causes the nonmonotonic $J_c(d)$. 

Now I turn to YBCO coated conductors, focusing on two main issues. 1. Can Eq. (\ref{jc}) account for the 
observed dependence $J_c(d)$ \cite{lanl1,lanl2}? 2. Can Eq. (\ref{jc}) help answer the question whether 
the observed $J_c(d)$ behavior is mostly determined by the thickness-dependent microstructure 
or the 2D pinning? Shown in Fig. 2 are the experimental data on the critical current 
$I_c=dJ_c(d)w$ of YBCO coated conductors on flexible metal tapes \cite{lanl2} along with the  
curve $I_c\propto\sqrt{d}$ which follows from Eq. (\ref{jc}) for $d\ll d_m$. For $d<2 \mu$m, the 2D pinning 
theory does capture the behavior of $I_c(d)$.  However, at larger d, the $I_c(d)$ values are weakly  
dependent on d, exhibiting a shallow minimum at thicknesses  $3<d<6 \mu$m which are just in the 2D-3D 
pinning crossover region. 

Although the 2D pinning model is in a reasonable agreement with the observed behavior of $J_c(d)$ at small d, the  
relative contribution of the 2D pinning as compared to the essential microstructural factors (lesser epitaxy and higher porosity 
in thicker films) at low fields remains unclear.  However, the 2D pinning contribution can be probed more directly by measuring 
$J_c(d)$ at higher B of the order of the irreversibility field $B^*$. If such measurements 
reveal the nonmonotonic dependence $J_c(d)$, it would unambiguously indicate the key role of the 2D pinning. Indeed, 
any thickness-dependent microstructure can hardly cause the maximum in $J_c(d)$ in macroscopic films with $d\sim 1\mu$m 
at high fields, while resulting in the decreasing $J_c(d)$ at lower fields. The nonmonotonic $J_c(d)$ can be observed if $d_m$ 
gets in the range $0.1<d<6\mu$m where $J_c(d)$ is usually measured. As follows from Eq. (\ref{dm}), this implies reducing 
$J_c$ from the typical zero field values $\sim 1-3 MA/cm^2$ down to $J_c\sim 10^4A/cm^2$. Such $J_c$ reduction can be achieved 
by increasing B and/or T (around 5T at 77K \cite{ornl}).  Another easily tested feature of the 2D pinning is that the crossover 
thickness (\ref{td}), above which $J_c$ becomes independent of d, increases as T and B are increased.   

The above results may also be applied to thin filaments of radius $d_m\ll R<L_c$ in a 
field $B\gg 2\pi J_cR/c$. For $B<B_s$, the critical current $I_c$ of the round 
filament can be obtained by averaging Eq. (\ref{jc}), $I_c=4\int_0^{R}yJ_c(2y)dx$, with 
the local "thickness" $2y=2\sqrt{R^2-x^2}$, and $\exp(-\sqrt{d_m/d})=1$. Hence, 
	\begin{equation}
	I_c=2.47c\sqrt{\gamma}R^{3/2}/\phi_0,
	\label{ic}
	\end{equation}
where $2.47=\sqrt{2}B(1/2,5/4)$, and $B(a,b)$ is the Euler beta function. The 2D pinning 
gives rise to a weaker size dependence of $I_c\propto R^{3/2}$ for $B<B_s$ and $I_c\propto R$ for $B>B_s$ 
(see Eq. (\ref{ehb1})) as compared to the 3D $I_c\propto R^2$. 

In summary, the 2D vortex pinning and self-field effects are suggested as possible mechanisms of the thickness dependence of 
$J_c$ in coated conductors. A nonmonotonic dependence of $J_c(d)$ is obtained which accounts both 
the 2D pinning and thermal fluctuations of vortices. An experimental test of the 2D pinning mechanism is proposed.  

This work was supported by the AFOSR MURI (F49620-01-1-0464), NSF  MRSEC (DMR 9214707), 
and DOE HEP (DE-FG02-91ER40643). 


\begin{figure}          
\caption{Thickness dependence of $J_c(d)$ described by Eq. (\ref{jc}). The 
inset shows a vortex a film, where gray dots are pinning centers in the interaction 
cylinder of radius $r_p$ and hight d.}
\label{fig.1}
\end{figure}

\begin{figure}          
\caption{Normalized critical currents of IBAD-on-metal substrates conductors for different 
YBCO thicknesses \cite{lanl2}. The thin solid line represents the $I_c$ data for YSZ 
crystalline substrates \cite{lanl1}. The bold solid line shows the fit to $I_c\propto\sqrt{d}$, 
as follows from Eq. (\ref{jc}) for $d\gg d_m$. }
\label{fig.2}
\end{figure}

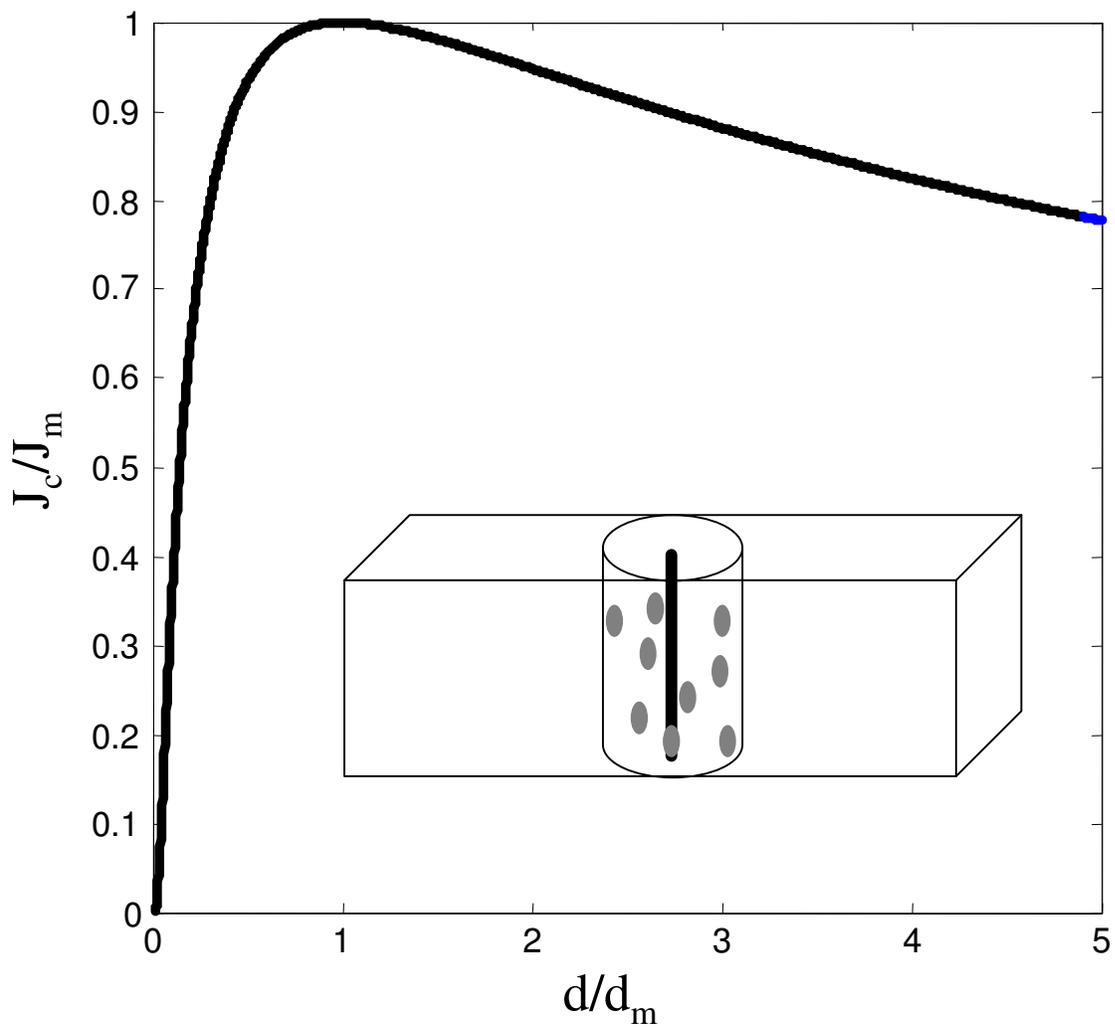
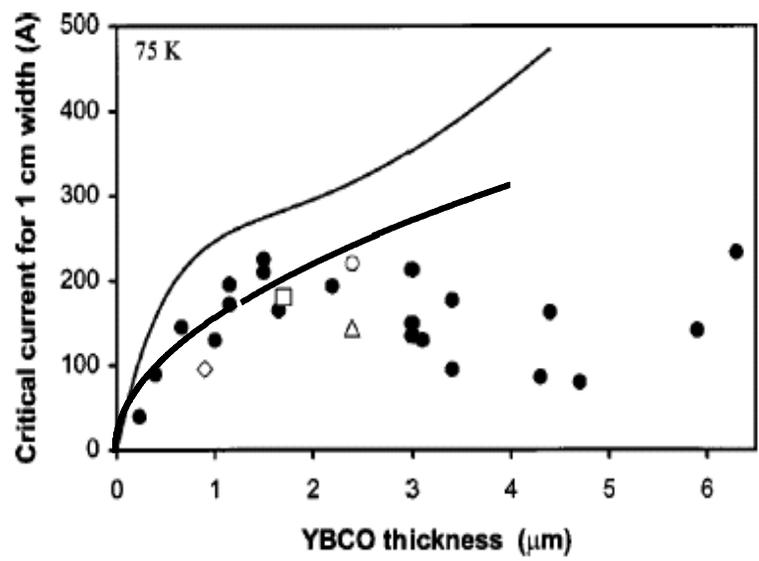
\end{document}